\documentstyle[aps,prb,amsfonts,twocolumn]{revtex}

\renewcommand{\vec}[1]{{\bf #1}}

\input{psfig}

\begin{document}

\draft

\title{Temporal fluctuations of waves
in weakly nonlinear disordered media$^1$}
\author{S. E. Skipetrov}
\address{Laboratoire de Physique et Mod\'elisation des Milieux
Condens\'es,\\ Universit\'e Joseph Fourier, Maison des Magist\`eres
--- CNRS,\\
B.P. 166, 38042 Grenoble Cedex 9, France\\
and\\
Department of Physics, Moscow State University, 119899 Moscow,
Russia}
\maketitle

\begin{abstract}
\begin{center}
{\bf Abstract} \end{center} We consider the multiple scattering of
a scalar wave in a disordered medium with a weak nonlinearity of
Kerr type. The perturbation theory, developed to calculate the
temporal autocorrelation function of scattered wave, fails at short
correlation times. A self-consistent calculation shows that for
nonlinearities exceeding a certain threshold value, the
multiple-scattering speckle pattern becomes unstable and exhibits
spontaneous fluctuations even in the absence of scatterer motion.
The instability is due to a distributed feedback in the system
``coherent wave + nonlinear disordered medium''. The feedback is
provided by the multiple scattering. The development of instability
is independent of the sign of nonlinearity.
\end{abstract}


\section{Introduction}
\label{intro}

Scattering\footnotetext[1]{Accepted for publication in {\em
Physical Review E.} You can contact the author at
http://www.ilc.msu.su/$^{\sim}$skipetr/ .} of waves in disordered
media has proved to be a nontrivial topic possessing intriguing and
still not completely understood
features.\cite{sheng95,papa97,gdr98,fouque98,vanrossum99}
Accordingly to the strength of disorder, one observes a variety of
propagation regimes ranging from ballistic transport, through
single scattering and wave diffusion, to the Anderson localization.
In this paper we are interested in the regime of wave diffusion,
corresponding to a relatively strong disorder, which is, however,
still largely insufficient to bring the system to the localization
transition ($k \ell \gg 1$, where $k$ is a wave number in the
medium, and $\ell$ is a mean free path).

It is well-known, that
multiple scattering of coherent wave in a disordered medium
results in a complicated spatial intensity
distribution $I(\vec{r}, t)$ known as a ``speckle pattern''.
The speckle pattern
is highly irregular and appears random to the eye.
It is now well established that the speckle pattern exhibits
large intensity fluctuations\cite{goodman76,ishimaru78,shapiro86}
$\left< \delta I(\vec{r}, t)^2 \right> \simeq \left< I(\vec{r}, t)
\right>^2$, originating from the interference of partial waves
arriving at $\vec{r}$ with completely randomized phases.
Here the angular brackets $\left< \cdots \right>$ denote ensemble
averaging, and $\delta I(\vec{r}, t) = I(\vec{r}, t) -
\left< I(\vec{r}, t) \right>$.
Besides, the speckle pattern possesses nontrivial
long-range spatial correlation
$C_{\delta I}(\vec{r}, \Delta \vec{r}) =
\left< \delta I(\vec{r}-\Delta\vec{r}/2, t)
\delta I(\vec{r} + \Delta\vec{r}/2, t) \right>$
even for $\Delta r > \ell$. This correlation is
due to interaction of
diffusing modes.\cite{zyuzin87,stephen87,pnini89,berk94}
If the points $\vec{r} \pm \Delta\vec{r}/2$ are far enough from the
boundaries of the medium, $C_{\delta I} \propto 1/\Delta r$.
Recently, it has been shown that in a particular case of a point
source of waves embedded inside a disordered medium,
there exists an infinite-range contribution to
$C_{\delta I}(\vec{r}, \Delta \vec{r})$
originating from scattering events taking place in the immediate
neighborhood of the source.\cite{shapiro99}
This contribution is highly sensitive to the short-distance properties
of disorder, as well as to the source size and shape.\cite{skip00a}

If the scatterers in the medium are allowed to move, $I(\vec{r}, t)$
fluctuates with time, and the statistics of these fluctuations
is also a subject of active research.
A wave propagating in a disordered, multiple-scattering medium
undergoes a large number of scattering events, and hence
the scattered intensity is highly sensitive to displacements
of scatterers.\cite{lee85,berk91,simons93} Consequently,
the decay of the intensity autocorrelation function
$C_{\delta I}(\vec{r}, \tau) =
\left< \delta I(\vec{r}, t)
\delta I(\vec{r}, t+\tau) \right>$
is considerably faster than in the single-scattering
case.\cite{gol84,maret87,stephen88}
Recently, long-range autocorrelation function of intensity
fluctuations has been measured,\cite{frank97}
and the existence of the universal conductance fluctuations
(analogous to that in disordered conductors) has been
demonstrated\cite{frank98} for optical waves.
Theoretical analysis of the temporal correlation function of
multiple-scattered waves has been extended to
amplifying disordered media,\cite{burkov97}
as well as to the case of intense incident waves producing
flows of scatterers in the disordered medium.\cite{skip98a,skip98b}
An additional contribution to $C_{\delta I}(\vec{r}, \tau)$,
originating from scattering in the immediate neighborhood of
source and/or detector, decaying much slower than all the previously
known contributions, is predicted to exist.\cite{skip00a}

High sensitivity of multiple-scattering speckle patterns
to scatterer motion gave rise to a new technique
for studying the scatterer dynamics in disordered, turbid
media, the so-called ``diffusing-wave
spectroscopy'' (DWS).\cite{maret87,pine88,pine89,pine90,weitz93}
The latter is now widely applied in concentrated colloidal
suspensions,\cite{maret87,pine88,pine89,pine90,weitz93,maret97}
foams,\cite{durian91,earnshow94,gopal95,hohler97,gopal97}
emulsions,\cite{gang95,hebraud97,lisy99}
granular\cite{menon97a,menon97b,durian00}
and biological\cite{boas97a,palmer99,lohwasser99} media.
Besides, the DWS has been extended to macroscopically
heterogeneous turbid media, providing a tool for imaging
of dynamic heterogeneities\cite{boas95,boas97b,heck97}
and visualization of scatterer flows\cite{heck97,skip98c,heck98}
in the bulk of the medium.
A generalization of DWS technique has been also accomplished for
anisotropic disordered media.\cite{stark96,stark97a,stark97b}
Recently, the DWS approach has
been extended to nonergodic turbid media.\cite{nisato00,scheffold00}

The above-mentioned, extensive studies of temporal fluctuations
of multiple-scattered waves, as well as the numerous
application of DWS, are all restricted to {\em linear\/} disordered media.
In general, little information is available on the subject of
multiple scattering in {\em nonlinear\/} disordered media.
Meanwhile, the question concerning the way in which the
nonlinearity affects the multiple-scattering speckle pattern
still remains open and continue to attract the researches.
Considerable efforts have been made to
understand the phenomena of coherent backscattering
in disordered media with
Kerr-type nonlinearity:\cite{agran91,agran93,heid95}
a narrow dip has been predicted to appear on the top of the
backscattering peak.
Weak localization effects are
shown to exist in the radiation of second harmonic and difference
frequency,\cite{agran93,agran88,agran89,kravtsov91}
though their experimental observation failed.\cite{yoo89}
Also studied with account for disorder is
the optical phase conjugation.\cite{kravtsov89,kravtsov90,yudson91,paass97}
More recently, correlations in transmission and reflection coefficients
of second harmonic waves have been investigated both theoretically
and experimentally,\cite{deboer93} and the angular correlation
functions of fundamental wave in a disordered medium with
Kerr-type nonlinearity have been calculated.\cite{bressoux00}
Despite the fact that theoretical description of wave scattering in nonlinear
media is complicated by the simultaneous presence of both
disorder and nonlinearity, the standard diagram technique for
impurity scattering has been extended to the case of disordered
medium with nonlinearity of Kerr type.\cite{wonderen92,wonderen94}

Very recently, it has been shown that the
speckle pattern resulting from the multiple scattering of coherent
wave in a {\em nonlinear\/} disordered medium with Kerr-type nonlinearity,
should be extremely
sensitive to changes of scattering potential,\cite{spivak00} i.e. much more
sensitive than the {\em linear\/} speckle pattern.
This high sensitivity has been explained by the multiplicity of
solutions of nonlinear wave equation.\cite{spivak00}
The multiplicity of solutions has been then shown to lead to
the temporal {\em instability\/} of the multiple-scattering
speckle pattern in nonlinear medium, resulting in {\em
spontaneous\/} fluctuations of scattered wave with time.\cite{skip00b}
An important prediction of Ref.\ \onlinecite{skip00b} is that
the nonlinearity should exceed some threshold value for the
instability to develop.
The threshold value is principally determined by the absorption length
$L_a$, or by the sample size $L$, if $L<L_a$.
The striking feature of the established result is that the
threshold value of nonlinearity tends to zero in an unbounded medium
without absorption. Purely elastic, unbounded nonlinear multiple-scattering
systems are therefore always unstable.
The physical origin of the instability is easy to understand.\cite{skip00b}
Nonlinearity modifies the phases of partial waves
propagating in the medium. The phase modifications are proportional
to the intensity $I(\vec{r}, t)$ and affect the mutual interference
of partial waves.
As it is this interference which is responsible for $I(\vec{r}, t)$,
a sort of feedback establishes in the medium.
A small modification of $I(\vec{r}, t)$ causes modifications of
phases of partial waves which, in their turn, produce changes of
$I(\vec{r}, t)$, and so on.
It is well-known that nonlinear wave systems with sufficiently strong,
positive feedback may become unstable.\cite{gibbs85,vor95}
As an example, we cite a family of nonlinear
optical systems with two-dimensional
feedback,\cite{vor95,aless91,arecchi99,ramazza96} where spontaneous
formation of complicated spatial structures is observed. Despite
the absence of disorder, such systems exhibit transition to
seemingly chaotic dynamics with
increasing nonlinearity.\cite{arecchi99,ramazza96}
An analogy can be drawn between the nonlinear
optical systems with two-dimensional
feedback and nonlinear disordered media by considering the scattering
as a (three-dimensional) feedback mechanism. In the case of
disordered media, however, the feedback is of random nature and
it is therefore hopeless to expect regular spatial structures to form.
Meanwhile, the instability can manifest itself in spontaneous
fluctuations of speckle pattern.
In order to clarify the issue of instability of speckle patterns
in nonlinear disordered media, we consider the following
questions:
\begin{itemize}

\item[({\em a\/}).]
Can the multiple scattering provide a positive feedback mechanism
for waves propagating in a nonlinear disordered medium?

\item[({\em b\/}).]
If ``yes'', how strong the nonlinearity should be for
the instability to develop?

\end{itemize}
A general announcement of our principal answers to the above
questions has been given in our recent Letter.\cite{skip00b} In the
present paper, we discuss and justify the assumptions and
approximations underlying our conclusions, provide the missing
details of calculations, and give a comprehensive discussion of
results. Also developed and discussed is the perturbation approach
to the calculation of the temporal autocorrelation function of
multiple-scattered wave in a nonlinear disordered medium. It is
important that the validity condition of the perturbation theory
coincides with the condition for the instability threshold as
obtained by using the self-consistent approach. In addition, we
give a detailed consideration to an experimentally important case
of moving scatterers, when the decrease of the time autocorrelation
function is due to a combined effect of spontaneous and
scatterer-motion-induced fluctuations of the speckle pattern.

The remainder of the paper is arranged as follows.
In Sec.\ \ref{nweq}, we introduce the nonlinear wave equation,
and discuss how the path integral approach can be applied for
its analysis. We also formulate the basic models and
approximations used throughout the paper.
Section \ref{linear} is devoted to linear disordered media.
In this section, we provide the expressions for the spatio-temporal
intensity correlation functions. Although correlations of
multiple-scattered waves in linear media are well
studied nowadays, we present their first, to our knowledge,
treatment with a simultaneous account for absorption,
boundary conditions at the sample surface, and scatterer motion.
The results of Sec.\ \ref{linear} serve as a base for
further calculations.
In Sec.\ \ref{dephasing}, we present a calculation of
dephasing of waves in a nonlinear disordered medium. Our
calculation takes into account the fluctuations of the local
refractive index due to nonlinear effects,
as well as the long-range spatial correlation of this fluctuations.
Three ``nonlinear'' contributions to the
dephasing are identified in addition to the usual, ``linear'' term
originating directly from the motion of scatterers.
Further, in Sec.\ \ref{perturb} we develop a perturbation theory
for calculation of the temporal autocorrelation function of a
multiple-scattered wave, and show its failure at short correlation times,
for sufficiently weak absorption. A condition of validity of
the perturbation theory is
established by comparing the linear and nonlinear contributions
to the dephasing found in Sec.\ \ref{dephasing}.
Section \ref{self} presents an alternative, self-consistent approach
to the calculation of the temporal autocorrelation of scattered wave.
Development of self-consistent theory requires some additional assumptions
which are also discussed in this section.
In Sec.\ \ref{disc}, the main results of our self-consistent
approach are presented and discussed. The multiple-scattering
speckle pattern is shown to exhibit spontaneous fluctuations even
in the absence of scatterer motion, which we interpret as a signature of
its instability. A comparison of self-consistent
and perturbative results is given, and the condition of
the speckle pattern instability is shown to coincide with the
condition of validity of the perturbation theory.
Finally, concluding remarks are presented in Sec.\ \ref{concl}.
In order to maintain the text of the paper readable,
we have chosen to collect the technical details of calculations
in three appendices. Appendix \ref{appa} is devoted to the
derivation of the field-field spatio-temporal correlation function.
In Appendix \ref{appb} we compute the spatio-temporal
long-range intensity correlation function. Appendix
\ref{appc} provides the details of calculations of path
distributions $\rho_s(\vec{r})$ and $\rho_s(\vec{r},\vec{r}^{\prime})$
defined in Sec.\ \ref{dephasing}.

\section{Wave equation and path integrals}
\label{nweq}

We consider a scalar monochromatic wave of frequency $\omega$
propagating in a random
medium with Kerr-type nonlinearity. The wave amplitude $\psi(\vec{r}, t)$
obeys a nonlinear wave equation:\cite{shen84,moloney91}
\begin{eqnarray}
&&\left\{ \nabla^2 + k_0^2 \left[ \varepsilon_0^{\prime} +i
\varepsilon_0^{\prime\prime} + \delta \varepsilon(\vec{r}, t) +
\varepsilon_2 \left| \psi(\vec{r}, t) \right|^2 \right] \right\}
\psi(\vec{r}, t) \nonumber \\
&&\hspace{4.4cm} = 0. \label{weq}
\end{eqnarray}
Here $k_0$ is the free-space wave number, $\varepsilon_0 =
\varepsilon_0^{\prime}
+i \varepsilon_0^{\prime\prime}$ is the
average (complex) dielectric function,
$\delta \varepsilon(\vec{r}, t)$ is the
fluctuating part of the dielectric function,
$\varepsilon_2$ is the nonlinear susceptibility\cite{optics}
(the two latter quantities are assumed to be real).
Eq.\ (\ref{weq}) is valid only if
$\delta \varepsilon(\vec{r}, t) + \varepsilon_2
\left| \psi(\vec{r}, t) \right|^2$ does not change significantly
on the time scale of $\omega^{-1}$.
The expression in the square brackets of Eq.\ (\ref{weq}) can
be considered as some ``effective'' dielectric function of the medium.
General analysis of Eq.\ (\ref{weq}) for arbitrary relation between
various terms comprising this function constitutes a formidable
task, and is not a purpose of this paper.
We assume the following hierarchy:
\begin{eqnarray}
\varepsilon_2 \left| \psi(\vec{r}, t) \right|^2 \ll
\delta \varepsilon(\vec{r}, t), \mbox{~~}
\varepsilon_2 \left| \psi(\vec{r}, t) \right|^2 \ll
\varepsilon_0^{\prime}, \mbox{~~}
\varepsilon_0^{\prime\prime} \ll \varepsilon_0^{\prime}.
\label{hierarchy}
\end{eqnarray}
In other words, we assume that the role of nonlinearity is less
significant than that of disorder, and that absorption is weak
allowing multiple scattering of waves in the medium.
It is then convenient to define the effective refractive index
$n_0 = (\varepsilon_0^{\prime})^{1/2}$, the absorption length
$\ell_a = n_0/(k_0 \varepsilon_0^{\prime\prime})$,
and the nonlinear coefficient
$n_2 = \varepsilon_2/(2 n_0)$, which determines the nonlinear
correction to the (linear) refractive index of the medium:
$n(\vec{r}, t) = n_0 + n_2 I(\vec{r}, t)$, where
$I(\vec{r}, t) = \left| \psi(\vec{r}, t) \right|^2$ is
the wave intensity.

In this paper, we study the fluctuations of the solution
of Eq.\ (\ref{weq}) with time $t$. In a linear medium
($\varepsilon_2 = 0$), these fluctuations can only be due to random
fluctuations of $\delta \varepsilon(\vec{r}, t)$ with time.
The fluctuations of $\psi(\vec{r}, t)$ are commonly characterized by the
autocorrelation function
$C_{\psi}(\vec{r}, \tau) = \left< \psi(\vec{r}, t)
\psi^*(\vec{r}, t+\tau) \right>$.
We assume that this autocorrelation
function is independent of $t$ which implies that for fixed $\vec{r}$,
$\psi(\vec{r}, t)$ represents a stationary
random process
(this is obviously true if the sample geometry and
the source distribution do not change with time, and if
$\delta \varepsilon(\vec{r}, t)$ is a stationary random process).
Me take $\delta \varepsilon(\vec{r}, t)$
to be a Gaussian random field
with zero mean and the correlation function
$C_{\delta \varepsilon}(\Delta\vec{r}, \tau) =
\left< \delta \varepsilon(\vec{r}-\Delta\vec{r}/2, t)
\delta \varepsilon(\vec{r}+\Delta\vec{r}/2, t+\tau) \right>$.
For a medium composed of point-like scatterers undergoing Brownian
motion with a diffusion coefficient $D_B$,\cite{gol84,stephen88}
\begin{eqnarray}
C_{\delta \varepsilon}(\Delta \vec{r}, \tau) =
\frac{4 \pi /(k^4 \ell)}{(4 \pi D_B \tau)^{3/2}}
\exp\left(
-\frac{\Delta r^2}{4 D_B \tau} \right),
\label{cde}
\end{eqnarray}
where $k = k_0 n_0$, and the mean free path $\ell \ll \ell_a$
is introduced (a weak scattering limit $k \ell \gg 1$ is assumed).
A natural time scale for scattering of waves in the medium
described by Eq.\ (\ref{cde}) is set by the
characteristic time needed for a scatterer to move a distance
of order of the wavelength:
$\tau_0 = (4 k^2 D_B)^{-1}$.
From here on, we will be interested in short correlation times
$\tau \ll \tau_0$.

In the linear case ($\varepsilon_2 = 0$), several approaches have
been elaborated to analyze Eq.\ (\ref{weq}).
We mention the diagrammatic
techniques,\cite{frisch68,rytov89}
theory of radiative transfer,\cite{ishimaru78}
and the method of path integrals.\cite{dashen79,gross83}
The three approaches are known to give equivalent results
for $C_{\psi}$ at $\tau \ll \tau_0$.
In the present paper, we adopt the method of path integrals that was
originally proposed
in the framework of quantum electrodynamics,\cite{feynman65} but later
has been successfully used in various areas of physics,\cite{schulman81}
and, in particular, for the analysis of wave scattering
problems.\cite{dashen79,gross83}
The method is based on the fact that the solution
$\psi(\vec{r}, t)$ of the wave equation
(\ref{weq}) can be written in a form of a functional integral,
with integration performed over all possible trajectories (paths)
going from the source of waves to $\vec{r}$.\cite{dashen79}
Since in the weak scattering limit
($k \ell \gg 1$) different trajectories can be considered
independent, it appears that the correlation function
$C_{\psi}$ reduces to the following
integral:\cite{pine88,pine89,pine90,weitz93}
\begin{eqnarray}
C_{\psi}(\vec{r}, \tau) = I_0 \int_{0}^{\infty} P(\vec{r},s)
\exp\left[ - \frac12 \left< \Delta \varphi^2(\tau) \right>_s
\right] d s, \label{path}
\end{eqnarray}
where $I_0$ is the average intensity in a nonabsorbing medium,
$P(\vec{r},s)$ is a weight coefficient of paths of length $s$,
and $\left< \Delta \varphi^2(\tau) \right>_s$ denotes the squared
phase difference
$\Delta \varphi(t, \tau) = \varphi(t+\tau) - \varphi(t)$,
averaged over various realizations of disorder, and over
all possible paths of the same length $s$.
From here on, we denote such an averaging by
$\left< \cdots \right>_s$.
Note that $\left< \Delta \varphi(\tau) \right>_s = 0$ for the
model of Brownian point-like scatterers.
Meanwhile,\cite{pine89,pine90,weitz93}
\begin{eqnarray}
\left< \Delta \varphi^2(\tau) \right>_s^{(0)} =
\frac{\tau}{\tau_0} \times \frac{s}{\ell},
\label{dphi0}
\end{eqnarray}
where the superscript $(0)$ denotes the linear case.
It is worth noting that
$\left< \Delta \varphi^2(\tau) \right>_s^{(0)}$ does not depend
neither on the sample geometry, nor on the source and detector
positions.
Its value is only determined by the scatterer dynamics
(through the single-scattering correlation time $\tau_0$),
and the path length $s$.
In contrast, $P(\vec{r},s)$ can only be calculated
if the sample geometry, source distribution, and detector
position $\vec{r}$ are specified.
In what follows, we restrict our analysis to a semi-infinite
medium occupying the half-space $z > 0$, and illuminated by a plane
monochromatic wave incident at $z = 0$.
For $s \gg \ell$,
$P(\vec{r},s)$ becomes\cite{pine89,pine90,weitz93,bicout93}
\begin{eqnarray}
P(\vec{r},s) =
\left( \frac{3 z^2}{4 \pi \ell s^3} \right)^{1/2}
\exp\left( -\frac{3 z^2}{4 s \ell}
-\frac{s}{\ell_a} \right).
\label{pn}
\end{eqnarray}
Once the field correlation
$C_{\psi}(\vec{r}, \tau)$ is known, the autocorrelation of
intensity fluctuations
$C_{\delta I}(\vec{r}, \tau) =
\left< \delta I(\vec{r}, t) \delta I(\vec{r}, t+\tau) \right>$
can be found applying the factorization approximation:
$C_{\delta I}(\vec{r}, \tau) =
| C_{\psi}(\vec{r}, \tau) |^2$.\cite{fact}

Combining together Eqs.\ (\ref{path})--(\ref{pn}), one obtains the
normalized autocorrelation function of multiple-scattered wave
in a semi-infinite disordered medium:
\begin{eqnarray}
g_1^{(L)}(\vec{r}, \tau) =
\frac{C_{\psi}(\vec{r}, \tau)}{C_{\psi}(\vec{r}, 0)} =
\exp\left\{ - \left[\alpha\left( \tau \right)
- \frac{\ell}{L_a} \right]
\frac{z}{\ell} \right\},
\label{glin}
\end{eqnarray}
where the superscript $(L)$ denotes the linear case,
$\alpha^2(\tau) = 3 \tau/(2 \tau_0) + \ell^2/L_a^2$,
and $L_a = (\ell \ell_a/3)^{1/2} \gg \ell$.
For the diffusely reflected wave, we assume
$z \simeq \ell$ and get
\begin{eqnarray}
g_1^{(L)}(\ell, \tau) \equiv g_1^{(L)}(\tau) =
\exp\left\{ - \alpha\left( \tau \right)
+ \frac{\ell}{L_a} \right\}.
\label{glin0}
\end{eqnarray}
From here on, we will use
$g_1(\vec{r}, \tau)$ to denote the normalized autocorrelation
function at a point $\vec{r}$ {\em inside\/} the medium, while
$g_1(\tau)$ --- for the normalized autocorrelation function
of {\em diffusely reflected\/} wave. The superscripts $(L)$ and
$(NL)$ will be used to distinguish between linear and nonlinear
cases.

Now we turn to the nonlinear medium.
Strictly speaking, the method of path integrals cannot be
applied for the analysis of Eq.\ (\ref{weq}), once
$\varepsilon_2 \neq 0$.
The failure of the path-integral technique follows from the
fact that this approach relies on the superposition principle
which is not valid for waves in nonlinear media.
However, if the nonlinearity is weak [which is ensured by
the first two inequalities of
Eq.\ (\ref{hierarchy})],
Eq.\ (\ref{path}) is still approximately valid provided that
its main ingredients $P(\vec{r}, s)$ and
$\left< \Delta \varphi^2(\tau) \right>_s$
are computed with account for nonlinear effects.
To simplify such a calculation, we assume that the nonlinearity
is sufficiently weak to validate the following two assumptions:

\begin{itemize}

\item[({\em i\/})]
Propagation of waves in a weakly
nonlinear disordered medium
is diffusive with a mean free path $\ell$ unaffected by the
nonlinearity.
This implies that nonlinear refraction is
negligible at distances of order
$\ell$, and consequently, that $\Delta n^2 k\ell \ll 1$,
where $\Delta n = n_2 I_0$,
and $I_0$ is the average intensity in the absence of absorption.
This assumption is an alternative
formulation of the fact that the role of nonlinearity is
much less significant than that of disorder [see also the first
two inequalities of Eq.\ (\ref{hierarchy})].

\item[({\em ii\/})]
Intensity of the third harmonic
remains always much smaller than the intensity of the
fundamental wave.
This implies either that $\psi(\vec{r}, t)$
is considered as a complex quantity [in this case, Eq.\ (\ref{weq})
is a nonlinear Schr\"{o}dinger equation,
$\left| \psi(\vec{r}, t) \right|^2$ is time-independent,
and the third harmonic is not generated at all],
or that the medium has a sufficient degree of dispersion
for the phase matching condition\cite{shen84,moloney91} to be violated:
$\left| \Delta k \right| \ell \gg 1$ with
$\Delta k = k_{3} - 3 k$ and
$k_{3}$ being the wave number at frequency $3 \omega$.

\end{itemize}

Assumption ({\em i\/}) allows us to consider the path
distribution $P(\vec{r}, s)$ being unaffected by nonlinearity.
The only object to be recalculated with account for
nonlinear effects is then
$\left< \Delta \varphi^2(\tau) \right>_s$.
Before going into an explicit calculation of
$\left< \Delta \varphi^2(\tau) \right>_s$, we devote the next section
to a brief derivation of some important results for linear medium.

\section{Correlations in a linear medium}
\label{linear}

As indicated above,
we consider a monochromatic plane wave incident
at the surface $z = 0$ of a semi-infinite medium occupying the
half-space $z > 0$.
The average intensity at $z > \ell$ can be then found in the
diffusion approximation:\cite{ishimaru78,karab99}
$\left< I(\vec{r}, t) \right> = I_0 \exp(-z/L_a)$.
The spatio-temporal correlation function of the field
is given by a solution of the Bethe-Salpeter equation
(see Appendix \ref{appa} for details of the calculation):
\begin{eqnarray}
C_{\psi}(\vec{r}, \Delta\vec{r}, \tau) &=&
\left< \psi(\vec{r}-\Delta\vec{r}/2, t)
\psi^*(\vec{r} + \Delta\vec{r}/2, t+\tau) \right>
\nonumber \\
&=&I_0
\frac{\sin(k \Delta r)}{k \Delta r}
\exp\left\{ -\frac{\Delta r}{2 \ell}
-\alpha\left( \tau \right) \frac{z}{\ell}
\right\}.
\label{tfield}
\end{eqnarray}

Let us now consider the correlation functions of intensity fluctuations.
In addition to $\delta I(\vec{r}, t)$, which is the deviation
of intensity from its average value,
it is convenient to define
$\Delta I(\vec{r}, t, \tau) = I(\vec{r}, t+\tau) - I(\vec{r}, t)$,
which is the change of the local intensity during the time interval
$\tau$. While
$\left< \delta I(\vec{r}, t) \right> =
\left< \Delta I(\vec{r}, t, \tau) \right> = 0$, the correlation functions
$C_{\delta I}(\vec{r}, \Delta\vec{r}, \tau)$
and
$C_{\Delta I}(\vec{r}, \Delta\vec{r}, \tau) =
\left< \Delta I(\vec{r}-\Delta\vec{r}/2, t, \tau)
\Delta I(\vec{r}+\Delta\vec{r}/2, t,
\tau) \right>$
for $\Delta r < \ell$
can be found in the factorization approximation:
\begin{eqnarray}
&&C_{\delta I}(\vec{r}, \Delta\vec{r}, \tau) =
\left|C_{\psi}(\vec{r}, \Delta\vec{r}, \tau) \right|^2,
\label{tint1}
\\
&&C_{\Delta I}(\vec{r}, \Delta\vec{r}, \tau) =
2 \left[ C_{\delta I}(\vec{r}, \Delta\vec{r}, 0) -
C_{\delta I}(\vec{r}, \Delta\vec{r}, \tau) \right].
\label{tint2}
\end{eqnarray}

Both correlation functions (\ref{tint1}) and (\ref{tint2}) decrease
exponentially with $\Delta r/ \ell$, and thus become negligible
for $\Delta r > \ell$. Intensity correlation persists, however, even
for two points separated by a distance $\Delta r > \ell$.
This correlation is due to the diffusive nature of wave transport in the
medium and can be found either using the Langevin
approach\cite{zyuzin87,pnini89}
or applying diagrammatic methods\cite{stephen87,berk94}.
We give the details of calculations in Appendix \ref{appb},
the final results are
\begin{eqnarray}
&&C_{\delta I}(\vec{r}, \Delta\vec{r}, \tau) = \frac{3}{(k \ell)^2}
I_0^2 \int_0^{\infty} d K\, K \nonumber \\
&&\times\, Q \left[ K, \sqrt{K^2 + \ell^2/L_a^2}, \frac{z}{\ell},
\frac{\Delta z}{\ell}, \alpha\left( \tau \right) \right] \nonumber
\\
&&\times\, J_0\left(K \Delta R/\ell \right), \label{tint3}
\\
&&C_{\Delta I}(\vec{r}, \Delta\vec{r}, \tau) = \frac{6}{(k \ell)^2}
I_0^2 \int_0^{\infty} d K\, K \nonumber \\
&&\times\, \Delta Q \left[K, \sqrt{K^2 + \ell^2/L_a^2},
\frac{z}{\ell}, \frac{\Delta z}{\ell}, \alpha\left( \tau \right),
\alpha(0) \right] \nonumber \\
&&\times\, J_0\left(K \Delta R/\ell \right). \label{tint4}
\end{eqnarray}
Here we use the cylindrical coordinates:
$\vec{r} = \left\{ \vec{R}, z \right\}$,
$J_0$ is the Bessel function of zeroth order, the function
$Q$ is defined in Appendix \ref{appb}, and
\begin{eqnarray}
\Delta Q \left[\cdots, \alpha\left( \tau \right), \alpha(0),
\right] &=& Q \left[\cdots, \alpha(0) \right] \nonumber \\
&-& Q \left[\cdots, \alpha\left( \tau \right) \right].
\label{dqdef}
\end{eqnarray}

Due to a rather complicated structure of the function $Q$
(see Appendix \ref{appb}), further calculations can be done only
approximately.
In the case of
$z \pm \Delta z/2, \Delta r \ll \ell/\alpha(\tau)$, we get
\begin{eqnarray}
C_{\delta I}(\vec{r}, \Delta\vec{r}, \tau) &\simeq& \frac34
\frac{I_0^2 }{(k \ell)^2} \frac{\ell}{z_g} \left[\frac{1}{\chi} -
\frac{1}{\sqrt{1+\chi^2}} \right] \nonumber \\
&\times& \exp\left[-2\alpha\left( \tau \right) \frac{z}{\ell}
\right], \label{lrs}
\end{eqnarray}
where $z_g$ is the geometrical average of
$z$-coordinates of the two points for which the correlation
is computed: $z_g = \sqrt{z^2 - \Delta z^2/4}$, and
$\chi = \Delta r /(2 z_g)$.
For $\alpha(\tau) = 0$, Eq.\ (\ref{lrs}) is exact.
If $\chi < 1$, the correlation behaves essentially as $1/\Delta r$, while
for $\chi > 1$ it becomes proportional to $z_g^2/\Delta r^3$.
For the correlation function of Eq.\ (\ref{tint4}) we find:
\begin{eqnarray}
&&C_{\Delta I}(\vec{r}, \Delta\vec{r}, \tau) \nonumber \\
&&\simeq \cases{ 4 \left[ \alpha\left(\tau \right) - \alpha(0)
\right] (z/\ell) C_{\delta I}(\vec{r}, \Delta\vec{r}, 0),\cr
\hspace{4cm} 2 \alpha(\tau) (z/\ell) \ll 1\cr 2 C_{\delta
I}(\vec{r}, \Delta\vec{r}, 0), \hspace{1.75cm} 2 \alpha(\tau)
(z/\ell) \gg 1} \label{didi}
\end{eqnarray}

\section{Dephasing of waves in a nonlinear medium}
\label{dephasing}

Consider a single wave path of length $s$ going from the source
of waves to some point $\vec{r}$.
The phase acquired by a wave traveling
along such a path can be written as
\begin{eqnarray}
\varphi(t) = \int_{0}^{s} k_0 n[\vec{r}(s_1), t]\, d s_1,
\label{ph}
\end{eqnarray}
where the integration is along the path, and
$n(\vec{r}, t) = n_0 + n_2 I(\vec{r}, t)$.
The squared difference $\Delta \varphi(t, \tau) =
\varphi(t+\tau) - \varphi(t)$, averaged over various realizations
of disorder, and over all possible paths
of length $s$, is found directly from Eq.\ (\ref{ph}):
\begin{eqnarray}
\left< \Delta \varphi^2(\tau) \right>_s =
\sum_{j=0}^3 \left< \Delta \varphi^2(\tau) \right>_s^{(j)},
\label{dphnl}
\end{eqnarray}
where the four contributions corresponding to $j = 0, \ldots, 3$
originate from different physical processes. Below, we give explicit
expressions of these terms and discuss their origin.

The first term in Eq.\ (\ref{dphnl}),
$\left< \Delta \varphi^2(\tau) \right>_s^{(0)}$ is the linear
term given by Eq.\ (\ref{dphi0}).
The next three terms, namely the terms
corresponding to $j = 1, 2, 3$, are absent in the linear case
and only appear because of nonlinear nature of wave interaction
with the medium. Explicit expressions for these terms are
\begin{eqnarray}
\left< \Delta \varphi^2(\tau) \right>_s^{(1)} &=& \left< \frac{2
n_2}{n_0 \ell} \frac{\tau}{\tau_0} \int_{0}^{s} \left< I(\vec{r})
\right> d s_1 \right>_s \nonumber \\
&=& \frac{2 n_2}{n_0} \frac{\tau}{\tau_0} \left<\left< I(\vec{r})
\right>\right>_s \frac{s}{\ell}, \label{dphi1}
\\
\left< \Delta \varphi^2(\tau) \right>_s^{(2)} &=& \left<
\frac{\pi}{n_0} k_0 n_2^2 \int_{0}^{s} C_{\Delta I}(\vec{r}, 0,
\tau)\, d s_1 \right>_s \nonumber \\
&=& \frac{\pi}{n_0}\, k_0 \ell\, n_2^2 \left< C_{\Delta I}(\vec{r},
0, \tau) \right>_s \frac{s}{\ell}, \label{dphi2}
\\
\left< \Delta \varphi^2(\tau) \right>_s^{(3)} &=& \left< k_0^2
n_2^2 \int_{0}^{s} \int_{0}^{s} C_{\Delta I}(\vec{r},
\Delta\vec{r}, \tau)\, d s_1\,  d s_2 \right>_s =
\nonumber \\
&=& (k_0 \ell)^2 n_2^2
\left< C_{\Delta I}(\vec{r}, \Delta\vec{r}, \tau) \right>_s
\left( \frac{s}{\ell} \right)^2.
\label{dphi3}
\end{eqnarray}
Here the integrations are assumed along wave paths of length
$s \gg \ell$ [in Eq.\ (\ref{dphi3}), both integrals are along the same
path]. Eq.\ (\ref{dphi2}) originates from the short-range correlation
of intensity fluctuations [see Eqs.\ (\ref{tfield})--(\ref{tint2})]:
\begin{eqnarray}
\left< \Delta \varphi^2(\tau) \right>_s^{(2)} &=& k_0^2 n_2^2
\left< C_{\Delta I}(\vec{r}, 0, \tau) \right>_s \int_0^s d s
\int_{-\ell}^{\ell} d (\Delta s) \nonumber \\
&&\left[ \frac{\sin(k \Delta s)}{k \Delta s} \right]^2
\exp\left(-\frac{\Delta s}{\ell} \right), \label{dphi2p}
\end{eqnarray}
where the wave path is assumed to be ballistic at distances
shorter than $\ell$. Eq.\ (\ref{dphi2p}) reduces to (\ref{dphi2})
for $k \ell \gg 1$.
Next, the term given by Eq.\ (\ref{dphi3})
is due to the long-range correlations of intensity
fluctuations [see Eq.\ (\ref{tint4})].
The averages entering into the right-hand sides of
Eqs.\ (\ref{dphi1})--(\ref{dphi2p}) are
\begin{eqnarray}
&&\left<\left< I(\vec{r}) \right>\right>_s = \int d^3 \vec{r}\,
\rho_s(\vec{r}) \left< I(\vec{r}) \right>, \label{im}
\\
&&\left< C_{\Delta I}(\vec{r}, 0, \tau) \right>_s = \int d^3
\vec{r}\, \rho_s(\vec{r})\, C_{\Delta I}(\vec{r}, 0, \tau),
\label{dim}
\\
&&\left< C_{\Delta I}(\vec{r}, \Delta\vec{r}, \tau) \right>_s =
\int d^3 \vec{r}_1 \int d^3 \vec{r}_2\; \rho_s(\vec{r}_1,
\vec{r}_2) \nonumber \\
&& \hspace{3cm} C_{\Delta I}(\vec{r}, \Delta\vec{r}, \tau),
\label{didim}
\end{eqnarray}
where the integrations are over the volume of the
disordered medium, and $\vec{r}_{1,2} = \vec{r} \pm \Delta\vec{r}/2$.
In Eqs.\ (\ref{im})--(\ref{didim}),
$\rho_s(\vec{r})$ is the probability density for a path
of length $s$, going from the source of waves to the point
$\vec{r}$,
to pass through the vicinity of $\vec{r}$,
and $\rho_s(\vec{r}_1, \vec{r}_2)$
is the probability density for the path to pass consequently
through the vicinities of $\vec{r}_1$ and $\vec{r}_2$.
These two ``path distributions'' should be calculated with account
for a particular geometry of disordered sample and source of waves.
Once the geometry is fixed, the calculation is straightforward.
For the case of a plane wave incident upon a semi-infinite
disordered medium, the calculations of
$\rho_s(\vec{r})$ and $\rho_s(\vec{r}_1, \vec{r}_2)$
are presented in Appendix \ref{appc}.

Let us discuss briefly the physical origins of nonlinear
contributions to the dephasing given by Eqs.\ (\ref{dphi1})--(\ref{dphi3}).
$\left< \Delta \varphi^2(\tau) \right>_s^{(1)}$ describes the change
of the effective wave number in the nonlinear medium:
$k = k_0 n_0 \rightarrow k(\vec{r}) =
k_0 [n_0 + n_2 \left< I(\vec{r}, t) \right>]$.
This contribution can be either positive or
negative, depending on the sign of $n_2$, but its absolute value
is always much smaller than
$\left< \Delta \varphi^2(\tau) \right>_s^{(0)}$,
as long as $\Delta n \ll n_0$.
$\left< \Delta \varphi^2(\tau) \right>_s^{(1)}$ can
therefore only cause a small
correction to the linear correlation function.
The next contribution, $\left< \Delta \varphi^2(\tau) \right>_s^{(2)}$,
originates from fluctuations of the local intensity, while
$\left< \Delta \varphi^2(\tau) \right>_s^{(3)}$ is due to the long-range
correlation of these fluctuations. An important difference
between the linear term (\ref{dphi0}), the first nonlinear term
(\ref{dphi1}), and the two last nonlinear terms (\ref{dphi2}), (\ref{dphi3}),
is that the latter terms do not depend explicitly on $\tau_0$.
The terms given by
Eqs.\ (\ref{dphi2}) and (\ref{dphi3}) are determined
by the {\em intensity fluctuations,} and not by the scatterer displacements.
This might seem to be a meaningless statement, as the intensity
fluctuations are, in their turn, caused by the scatterer motion.
The important point is that the scatterer motion is not the only
possible reason for the fluctuations of intensity with time.
Weak, spontaneous fluctuations of
$I(\vec{r}, t)$ (due to thermal fluctuations of various
parameters, vibrations in the  experimental setup, fluctuations
of the incident wave, etc.) are inevitable in real physical systems.
Eqs.\ (\ref{dphi2}) and (\ref{dphi3}) provide a mechanism for this
spontaneous and generally weak fluctuations to affect
the dephasing $\left< \Delta \varphi^2(\tau) \right>_s$ and,
consequently, the temporal correlation function of scattered wave.

\section{Perturbation theory}
\label{perturb}

As stated in the title,
the present paper is devoted to {\em weakly\/}
nonlinear disordered media. We limit ourselves to a weak
nonlinearity, as otherwise the problem becomes too involved.
Above, we have already mentioned that we assume
$\Delta n^2 k \ell \ll 1$,
and that this condition allows us to consider
the transport of average intensity to remain unaffected by
the nonlinearity [assumption ({\em i\/}) of Sec.\ \ref{nweq}].
This allows us to use ``linear'' results,
$\left< I(\vec{r}) \right> =
I_0 \exp(-z/L_a)$ and Eqs.\ (\ref{rhoa}), (\ref{rhob})
of Appendix \ref{appc},
for $\left< I(\vec{r}) \right>$,
$\rho_s(\vec{r})$, and $\rho_s(\vec{r}_1, \vec{r}_2)$ in
Eqs. (\ref{im})--(\ref{didim}).
It seems then natural to assume that
$C_{\Delta I}(\vec{r}, 0, \tau)$ and
$C_{\Delta I}(\vec{r}, \Delta\vec{r}, \tau)$
are also close to their linear values.
We can therefore replace these correlation functions
in Eqs.\ (\ref{dim}) and (\ref{didim}) by the expressions
found in Sec.\ \ref{linear}.
Then, making use of Eqs.\ (\ref{tint1})--(\ref{tint4}),  and
performing necessary integrations, we obtain from
Eqs. (\ref{dphi1})--(\ref{dphi3}):
\begin{eqnarray}
\left< \Delta\varphi^2(\tau) \right>_s^{(1)} &=& 2 \Delta n
\frac{\tau}{\tau_0} \left\{ 1 - H\left[ \alpha(0) \sqrt{\frac{s}{12
\ell}} \;\right] \right\} \frac{s}{\ell},
\label{dphi11} \\
\left< \Delta\varphi^2(\tau) \right>_s^{(2)} &=& 2 \pi k_0 \ell
\Delta n^2 \left\{ H\left[ \alpha(\tau) \sqrt{\frac{s}{3 \ell}}
\;\right] \right. \nonumber \\
&-& \left. H\left[ \alpha(0) \sqrt{\frac{s}{3 \ell}} \;\right]
\right\} \frac{s}{\ell},
\label{dphi22}\\
\left< \Delta\varphi^2(\tau) \right>_s^{(3)} &=& 6 \Delta n^2
S\left[ \alpha(\tau)\sqrt{s/\ell}, \alpha(0)\sqrt{s/\ell}\; \right]
\nonumber \\
&\times& \left(\frac{s}{\ell}\right)^{3/2}. \label{dphi33}
\end{eqnarray}
Here $H(x) = \sqrt{\pi} x \exp(x^2) [1-{\rm Erf}(x)]$ and $n_0 = 1$
is assumed for simplicity. The function $S(u, v)$ in Eq.\
(\ref{dphi33}) is
\begin{eqnarray}
S(u, v) &=& 9 \int_0^{\infty} d R\, R \int_0^{\infty} d K\, K
\int_0^{\infty} d z \int_0^{2 z} d (\Delta z) \nonumber \\
&\times& f(z, \Delta z, R) \Delta Q(K, \sqrt{K^2 + v^2}, z, \Delta
z, u, v) \nonumber \\
&\times& J_0(KR), \label{s}
\\
f(z, \Delta z, R) &=& \frac{2 z + \sqrt{ R^2 + \Delta z^2}}{\sqrt{
R^2 + \Delta z^2}} \nonumber \\
&\times& \exp\left[ -\frac{3}{4}\left(2 z + \sqrt{ R^2 + \Delta
z^2} \right)^2 \right] \nonumber
\\
&-& \frac{2 z + \sqrt{ R^2 + 4 z^2}}{\sqrt{ R^2 + 4 z^2}} \nonumber
\\
&\times& \exp\left[ -\frac{3}{4}\left(2 z + \sqrt{ R^2 + 4 z^2}
\right)^2 \right]. \label{f}
\end{eqnarray}
Unfortunately, integrations in Eq.\ (\ref{s}) cannot be
performed in the general case. We find, however, the following
approximate results:
\begin{eqnarray}
S(u, v) \simeq \cases{
(u-v),
&$u - v \leq 1, v \leq 1$, \cr
1,
&$u - v > 1, v \leq 1$, \cr
(u-v) v^{-3},
&$u - v \leq 1, v > 1$, \cr
v^{-3},
&$u - v > 1, v > 1$.}
\label{sa}
\end{eqnarray}
Here numerical factors of order unity are omitted before each
of the four asymptotic expressions.

\begin{figure}[t]
\vspace{0.5cm} \psfig{file=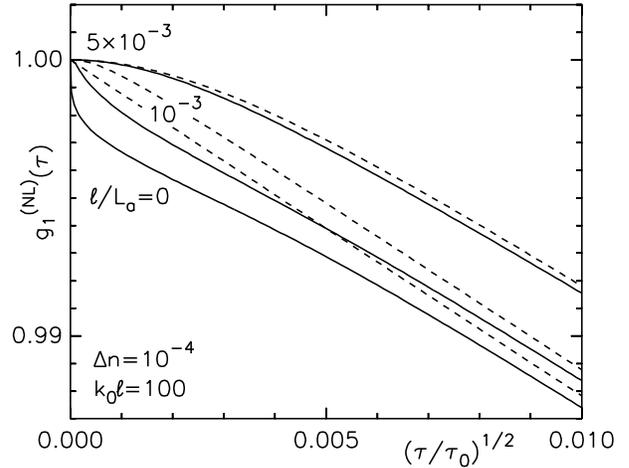,width=8cm} \vspace{0.5cm}
\caption{Normalized temporal autocorrelation function of a wave
diffusely reflected from a semi-infinite nonlinear disordered
medium, calculated using perturbation theory for $\Delta n =
10^{-4}$, $k_0 \ell = 100$, and the values of $\ell/L_a$ indicated
near each curve (solid lines). Dashed lines show corresponding
results for a linear medium ($\Delta n = 0$).} \label{fig1}
\end{figure}

While approximate, the above results enable one to compute the
temporal autocorrelation function $g_1^{(NL)}(\tau)$ of diffusely
reflected wave
numerically using Eqs.\ (\ref{path}), (\ref{pn}),
and (\ref{dphnl}):
\begin{eqnarray}
&&g_1^{(NL)}(\tau) = W\left[
\left< \Delta \varphi^2(\tau) \right>_s \right]/W[0],
\label{g1nl}
\\
&&W\left[ \left< \Delta \varphi^2(\tau) \right>_s \right] \nonumber
\\
&& = \int_{0}^{\infty} P(\ell,s) \exp\left[ - \frac12 \left< \Delta
\varphi^2(\tau) \right>_s \right] d s. \label{pathnl}
\end{eqnarray}
We present the results of the calculation
in Fig.\ \ref{fig1} for fixed $\Delta n$, $k_0 \ell$, and the
values of the inverse absorption length $\ell/L_a$ indicated
near each curve.
The ``linear'' correlation functions, corresponding to the
same three values of the absorption length, and to $\Delta n = 0$,
are shown by dashed lines.
For weak absorption ($\ell/L_a = 0$, $10^{-3}$),
the initial (i.e. at short correlation delay times $\tau$)
decrease of the ``nonlinear'' autocorrelation function is
much faster than that of the linear one.
Hence, our perturbation approach fails at short times.
Indeed, the results of this section are based on the assumption
that
$C_{\Delta I}(\vec{r}, 0, \tau)$ and
$C_{\Delta I}(\vec{r}, \Delta\vec{r}, \tau)$
in the nonlinear medium are close to their values in
the linear one. Applying the factorization approximation,
we find that this implies that
$1 - | g_1^{(NL)}(\tau) |^2 \approx
1 - | g_1^{(L)}(\tau) |^2$.
Although this condition seems to hold well for sufficiently
large $\tau$, it can be violated at small $\tau$, where, as follows
from Fig.\ \ref{fig1}, $1 - | g_1^{(NL)}(\tau) |^2$
can become much
larger than $1 - | g_1^{(L)}(\tau) |^2$,
if absorption is sufficiently weak.

To estimate the region of validity of our perturbation approach, we
require that the linear contribution to the dephasing $\left<
\Delta\varphi^2(\tau) \right>_s$ given by Eq.\ (\ref{dphi0}),
should be considerably greater than the sum of nonlinear
contributions [Eqs.\ (\ref{dphi11})--(\ref{dphi33})]. As the
longest path length contributing to the integral of Eqs.\
(\ref{path}) and (\ref{pathnl}) is $s \sim \ell/\alpha^2$, we
obtain the conditions of validity of the perturbation theory in the
form:
\begin{eqnarray}
\Delta n^2 \alpha(\tau)^{-2}
\left[ k_0 \ell + \alpha(\tau)^{-1} \right]
\ll 1.
\label{validity}
\end{eqnarray}

Since $\alpha(\tau)$ is an increasing function of its argument,
and $\alpha(0) = \ell/L_a$, condition
$\Delta n^2 (L_a/\ell)^2 [k_0 \ell + L_a/\ell] \ll 1$
ensures Eq.\ (\ref{validity}) at any $\tau$.
It is the case for the upper curve of Fig.\ \ref{fig1},
corresponding to $\ell/L_a = 5 \times 10^{-3}$. For such a strong absorption,
the perturbation theory is valid at any $\tau$, and the nonlinear
autocorrelation function is close to the linear result.
By contrast, if $\Delta n^2 (L_a/\ell)^2 [k_0 \ell + L_a/\ell]
> 1$, the perturbation approach can be applied only for sufficiently
long correlation times
$\tau > \tau_c$, where the critical time $\tau_c$ is determined
by Eq.\ (\ref{validity}). For the lower curve of Fig.\ \ref{fig1},
corresponding to $\ell/L_a = 0$, we find
$(\tau_c/\tau_0)^{1/2} \approx 2 \times 10^{-3}$.
It is worthwhile to note that in the absence of absorption,
even an infinitely small nonlinearity ($\Delta n \rightarrow 0$)
suffices for the perturbation theory to fail at
$\tau/\tau_0 \rightarrow 0$.
Our condition of validity of the perturbation approach
(\ref{validity}) is consistent with the result
of Spivak and Zyuzin,\cite{spivak00}
who have shown that the perturbation analysis
of the sensitivity of speckle
pattern in a nonlinear disordered medium
to changes of scattering potential
fails for $\Delta n^2 (L/\ell)^3 > 1$, where $L \gg \ell$ is
the typical size of the medium.

\section{Self-consistent analysis}
\label{self}

As demonstrated in the previous section, the perturbation theory
fails to describe the temporal autocorrelation function
of wave diffusely reflected from a nonlinear medium for
$\tau < \tau_c$, where $\tau_c$ is defined by Eq.\ (\ref{validity}).
To calculate $g_1^{(NL)}(\tau)$ at such short times,
one has to use a non-perturbative, self-consistent analysis.
However, the possibility of performing such an analysis is
considerably limited by the mathematical
complexity of the considered problem.
To make the self-consistent analysis possible, we adopt the following
two additional assumptions:

\begin{itemize}

\item[({\em iii\/})]
The statistics of a wave field scattered in
a weakly nonlinear disordered medium, is close to Gaussian.
Consequently, the factorization approximation
holds in a weakly nonlinear medium:
$C_{\delta I}(\vec{r}, \tau) \simeq | C_{\psi}(\vec{r}, \tau) |^2$.

\item[({\em iv\/})]
The functional form of the $\vec{r}$-dependence
of $g_1^{(NL)}(\vec{r}, \tau)$ is the same as that of
$g_1^{(L)}(\vec{r}, \tau)$:
$g_1^{(NL)}(\vec{r}, \tau) = \exp[ - \beta(\tau)\, z/\ell ]$,
where $\beta(\tau)$ is some unknown function, which can depend
not only on $\tau$ but also on other parameters of
the problem (namely, on $\ell/L_a$, $k_0 \ell$, $\Delta n$).

\end{itemize}

Strictly speaking,
the above assumptions define a sort of perturbation theory,
but now we do not limit the values of deviations of intensity
and field correlation functions from their values in the
linear case.
Instead, we assume that the nonlinearity does not cause
significant modifications of the {\em statistics\/} of scattered waves
[assumption ({\em iii\/})] and of the
{\em functional form\/} of the field correlation function
[assumption ({\em iv\/})].
Note that now the autocorrelation function $g_1^{(NL)}(\vec{r}, \tau)$
can deviate significantly from $g_1^{(L)}(\vec{r}, \tau)$.
Condition ({\em iv\/}) fixes the functional form of this
deviation, but implies no constraints on its absolute value.

Obviously, both the assumptions ({\em iii\/}) and ({\em iv\/})
require the nonlinearity to be weak.
Assumption ({\em iii\/}) is justified under
the same conditions as ({\em i\/}) (see Sec.\ \ref{nweq}),
since the Gaussian statistics
of the {\em total\/} scattered wave field $\psi(\vec{r}, t)$
is a consequence of the complete randomization of phases of
{\em partial\/} waves arriving at $\vec{r}$.
The reason for the randomization is that the typical
distance $\ell$ between individual scattering events in
a multiple-scattering sequence is much larger than the wavelength
($k \ell \gg 1$).\cite{shapiro86}
Obviously, such a mechanism of phase randomization is equally
effective in both linear and weakly nonlinear media,
as long as ({\em i\/}) holds.

To justify the ansatz of assumption ({\em iv\/}),
we apply Eq.\ (\ref{path}) and write $g_1^{(NL)}$ as
\begin{eqnarray}
g_1^{(NL)}(\vec{r}, \tau) &\approx& \left( \frac{3 z^2}{4 \pi \ell}
\right)^{1/2} \exp\left( \frac{z}{L_a} \right) \nonumber \\
&\times& \int_{0}^{s_1(\tau)} \frac{1}{s^{3/2}} \exp\left( -\frac{3
z^2}{4 \ell s} \right) ds, \label{g1ps}
\end{eqnarray}
where we assumed that
$s/\ell_a + (1/2)\left< \Delta \varphi^2(\tau) \right>_s$ increases
monotonically with $s$ and becomes of order
unity at $s = s_1(\tau) \gg s_0 = z^2/(2 \ell)$.
After performing integration, Eq.\ (\ref{g1ps}) can be approximately
rewritten as
$\exp(z/L_a)\{1 - z [3/(\pi \ell s_1)]^{1/2}\} \simeq
\exp[-\beta(\tau)\, z/\ell]$ with
$\beta z/\ell \simeq z (\ell s_1)^{-1/2} - z/L_a \ll 1$.
In the opposite limit of $\beta z / \ell \gg 1$, $g_1^{(NL)}(\vec{r}, \tau)$
vanishes and the functional form of its $z$-dependence is of no importance.
Anyway, it will be seen from the following that the exact functional form
of the $\vec{r}$-dependence of $g_1(\vec{r}, \tau)$
is not of crucial importance, since $g_1$ is
integrated over the whole medium during the calculation of
the correlation function of diffusely reflected wave.

Making use of Eqs.\ (\ref{dphi1})--(\ref{dphi3}) and relying on
the assumption ({\em iii\/}), ({\em iv\/}), we recalculate the
nonlinear contributions to the dephasing
$\left< \Delta \varphi^2(\tau) \right>_s$. Due to the assumption
({\em iv\/}), the result can be obtained from
Eqs.\ (\ref{dphi11})--(\ref{dphi33}) by a simple substitution:
$\alpha(\tau) \rightarrow \beta(\tau) + \alpha(0)$.
Recalling that $\alpha(0) = \ell/L_a$, we obtain:
\begin{eqnarray}
\left< \Delta\varphi^2(\tau) \right>_s^{(1)} &=& 2 \Delta n
\frac{\tau}{\tau_0} \left[ 1 - H\left( \frac{\ell}{L_a}
\sqrt{\frac{s}{12 \ell}} \;\right) \right] \frac{s}{\ell},
\label{d11} \\
\left< \Delta\varphi^2(\tau) \right>_s^{(2)} &=& 2 \pi k_0 \ell
\Delta n^2 \left\{ H\left[ \left(\beta + \frac{\ell}{L_a} \right)
\sqrt{\frac{s}{3 \ell}} \;\right] \right. \nonumber \\
&-& \left. H\left[ \frac{\ell}{L_a} \sqrt{\frac{s}{3 \ell}}
\;\right] \right\} \frac{s}{\ell},
\label{d22} \\
\left< \Delta\varphi^2(\tau) \right>_s^{(3)} &\simeq& 6 \Delta n^2
\nonumber \\
&\times& \cases{ \beta \times (s/\ell)^2, \cr \hspace{5mm} \beta
\sqrt{s/\ell} \leq 1, (\ell/L_a) \sqrt{s/\ell} \leq 1\cr
(s/\ell)^{3/2}, \cr \hspace{5mm} \beta \sqrt{s/\ell} > 1,
(\ell/L_a) \sqrt{s/\ell} \leq 1 \cr \beta \times (L_a/\ell)^3
\times \sqrt{s/\ell}, \cr \hspace{5mm} \beta \sqrt{s/\ell} \leq 1,
(\ell/L_a) \sqrt{s/\ell} > 1 \cr (L_a/\ell)^3, \cr \hspace{5mm}
\beta \sqrt{s/\ell} > 1, (\ell/L_a) \sqrt{s/\ell}
> 1} \label{d33}
\end{eqnarray}

As follows from Eqs.\ (\ref{d11})--(\ref{d33}), the nonlinear dephasing
depends on the unknown function $\beta$.
Since the temporal autocorrelation function of diffusely reflected wave
$g_1^{(NL)}(\tau) = \exp[-\beta(\tau)]$ is determined by
$\beta$ as well, Eqs.\ (\ref{g1nl}) and (\ref{pathnl})
allow us to formulate a self-consistent equation for $\beta$:
\begin{eqnarray}
&&\exp\left[-\beta\left( \tau \right) \right] =
F\left[ \beta\left( \tau \right) \right],
\label{beq}
\end{eqnarray}
where
$F(\beta) = W\left[\left< \Delta \varphi^2(\tau) \right>_s \right]/W[0]$,
$W[\cdots]$ is defined by Eq.\ (\ref{pathnl}), and
$\left< \Delta \varphi^2(\tau) \right>_s$ is a sum of terms given
by Eqs.\ (\ref{dphi0}), (\ref{d11})--(\ref{d33}).
Eq.\ (\ref{beq}) is the main result of our self-consistent analysis.
Although the functional form of $F(\beta)$ is rather complicated,
and Eq.\ (\ref{beq}) cannot be solved analytically, the numerical
solution is straightforward and can be carried out for any values
of $\Delta n$, $\ell/L_a$, $k_0 \ell$ and $\tau$.
Eq.\ (\ref{beq}) can be considered as a self-consistent equation
for the autocorrelation function of diffusely reflected
wave, since $\beta(\tau)$ and $g_1^{(NL)}(\tau)$ are
directly related. It is worthwhile to note that in the
absence of nonlinearity ($\Delta n = 0$), Eq.\ (\ref{beq})
yields $\beta(\tau) = \alpha(\tau) - \ell/L_a$, and hence
Eq.\ (\ref{glin0}) is recovered for $g_1^{(L)}(\tau)$.

\section{Results and discussion}
\label{disc}

\begin{figure}[t]
\vspace{0.5cm} \psfig{file=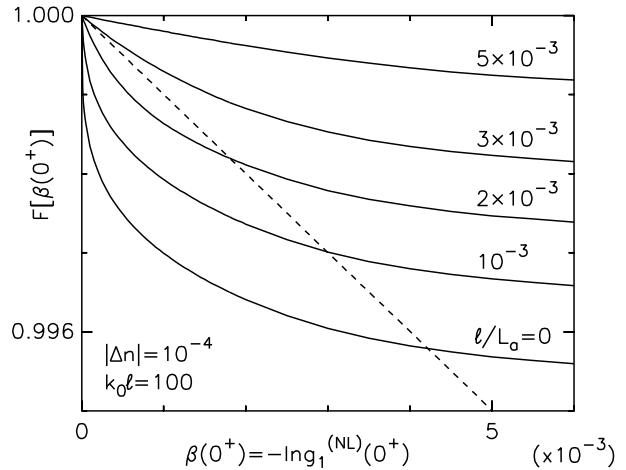,width=8cm} \vspace{0.5cm}
\caption{Graphical solution of Eq.\ (\ref{beq}) at $\tau/\tau_0
\rightarrow 0$. Solid lines show $F[\beta(0^+)]$ for $\left| \Delta
n \right| = 10^{-4}$, $k_0 \ell = 100$, and the values of
$\ell/L_a$ indicated near each curve. Dashed line is
$\exp[-\beta(0^+)]$. If absorption is weak ($\ell/L_a = 0$,
$10^{-3}$, and $2 \times 10^{-3}$), Eq.\ (\ref{beq}) has two
solutions $\beta(0^+) = 0$ and $\beta(0^+) > 0$, while for strong
absorption ($\ell/L_a = 3 \times 10^{-3}$ and $5 \times 10^{-3}$),
the second solution disappears.} \label{fig2}
\end{figure}

We start the analysis of Eq.\ (\ref{beq}) from the case of immobile
scatterers, taking a limit of $\tau/\tau_0 \rightarrow 0$. We
denote the autocorrelation functions corresponding to this limit by
$g_1^{(L), (NL)}(0^+)$ in order to distinguish them from $g_1^{(L),
(NL)}(0)$, which correspond to $\tau = 0$.\cite{unstable1}
Obviously, $g_1^{(L)}(0^+) = g_1^{(L)}(0) = 1$. In a nonlinear
medium, Eqs.\ (\ref{d22}) and (\ref{d33}) can still contribute to
the dephasing even for $\tau/\tau_0 \rightarrow 0$. These
contributions are insensitive to the sign of $\Delta n$. A
corresponding value of $\beta(0^+)$ and, consequently, of the field
autocorrelation function $g_1^{(NL)}(0^+)$ can be found by solving
Eq.\ (\ref{beq}) numerically. In Fig.\ \ref{fig2}, we plot the
left-hand and the right-hand sides of Eq.\ (\ref{beq}) for fixed
$\left| \Delta n \right| = 10^{-4}$, $k_0 \ell = 100$, and for
several values of $\ell/L_a$. If absorption is strong ($\ell/L_a
\gtrsim 3 \times 10^{-3}$ for considered $\left| \Delta n \right|$
and $k_0 \ell$), Eq.\ (\ref{beq}) has a unique solution $\beta(0^+)
= 0$ which corresponds to $g_1^{(NL)}(0^+) = 1$. However, a second
solution $\beta(0^+) > 0$ appears for sufficiently weak absorption
($\ell/L_a < 3 \times 10^{-3}$). The point of appearance of the
second solution is a {\em bifurcation\/} point of Eq.\ (\ref{beq}).
To choose the solution realizable in a physical system, we note
that the first solution [$\beta(0^+) = 0$] exists only for
$\tau/\tau_0 = 0$, and disappears at finite $\tau/\tau_0$, since
$F(0) < 1$ for $\tau/\tau_0 > 0$. This solution is therefore
inaccessible by continuity, and ``unstable'' with respect to small
scatterer displacements. The physically realizable solution should
represent the limit of $\beta(\tau)$ for $\tau/\tau_0 \rightarrow
0$, which is given by the second solution of Eq.\ (\ref{beq}). It
is therefore this solution which one expects to be realized in a
real physical system.

The fact that Eq.\ (\ref{beq}) can have a positive solution
$\beta(0^+) > 0$ is a very important issue, since $\beta(0^+) > 0$
leads to $g_1^{(NL)}(0^+) = \exp[-\beta(0^+)] < 1$. A value of the
temporal autocorrelation function which is less than unity is
commonly associated with temporal fluctuations of scattered waves.
In the considered case, however, the reason for these fluctuations
is not the motion of scatterers, as the limit of $\tau/\tau_0
\rightarrow 0$ corresponds to immobile scatterers. The fluctuations
are {\em spontaneous\/} and represent a clear signature of {\em
instability\/} of the multiple-scattering speckle
pattern.\cite{unstable2}

Despite a rather complicated structure of the function $F(\beta)$
in Eq.\ (\ref{beq}), a relation between the parameters of the problem
corresponding to the onset of the speckle pattern instability
can be found analytically.
As follows from Fig.\ \ref{fig2}, the initial [at $\beta(0^+) = 0$]
decay of $F[\beta(0^+)]$ should be faster than the decay of
$\exp[-\beta(0^+)]$, for the second solution of Eq.\ (\ref{beq})
to appear. A surface in a three-dimensional space of
the problem parameters $\Delta n$, $\ell/L_a$, $k_0 \ell$,
separating the stable [$\beta(0^+) = 0$] and unstable [$\beta(0^+) > 0$]
regions, is therefore given by the equation
\begin{eqnarray}
\frac{\partial}{\partial \beta(0^+)}
F[\beta(0^+)] \bigg|_{\beta(0^+) = 0} = -1.
\label{surface}
\end{eqnarray}

Recalling that $L_a/\ell \gg 1$, $k_0 \ell \gg 1$ is assumed,
we obtain from Eq.\ (\ref{surface}):
\begin{eqnarray}
p =
\Delta n^2 \left(\frac{L_a}{\ell} \right)^2
\left[  k_0 \ell  + \frac{L_a}{\ell} \right] \simeq 1,
\label{stbl}
\end{eqnarray}
where we introduce a control parameter $p$, and a numerical factor
of order unity is omitted. If $p<1$, the multiple-scattering
speckle pattern is stable [$g_1^{(NL)}(0^+) = 1$], while for $p>1$,
an instability shows up leading to $g_1^{(NL)}(0^+) < 1$. A
striking feature of Eq.\ (\ref{stbl}) is that $p$ can become larger
than unity even for very small $\left| \Delta n \right|$, provided
that the extensive parameter $L_a/\ell$ is large enough. Our
condition of the speckle pattern stability $p<1$ agrees with the
condition of validity of the perturbation theory developed in Sec.\
\ref{perturb} [Eq.\ (\ref{validity})], evaluated at $\tau/\tau_0 =
0$. This readily explains the failure of the perturbation theory
for short correlation times and weak absorption: perturbation
approach is not suitable for description of unstable regimes.
Moreover, the fact that the condition of validity of the
perturbation theory and the speckle pattern stability condition
agrees indicates that the additional assumptions ({\em iii\/}) and
({\em iv\/}) of Sec.\ \ref{self} are not essential for obtaining
unstable regimes.

\begin{figure}
\vspace{0.5cm} \psfig{file=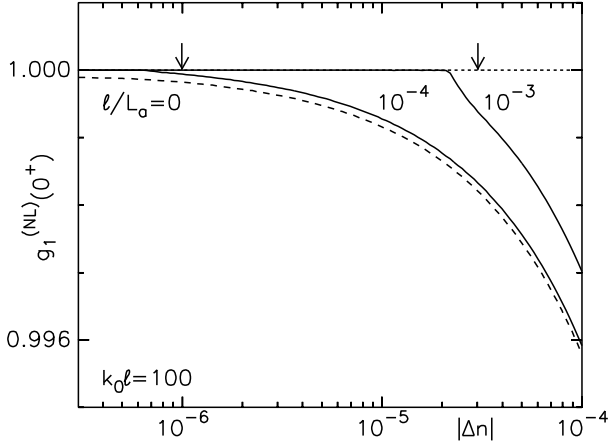,width=8cm} \vspace{0.5cm}
\caption{``Bifurcation diagram'' for a wave scattered in a
semi-infinite nonlinear disordered medium. Solid lines show the
temporal autocorrelation function of diffusely reflected wave at
$\tau/\tau_0 \rightarrow 0$ (immobile scatterers) for $k_0 \ell =
100$ and the values of $\ell/L_a$ indicated near each curve (dashed
line is for $\ell/L_a = 0$). The threshold values of $\left| \Delta
n \right|$ following from Eq.\ (\ref{stbl}) are shown by arrows.
$g_1^{(NL)}(0^+) < 1$ corresponds to spontaneous fluctuations of
scattered wave, which is a manifestation of the speckle pattern
instability. Dotted line is the linear result $g_1^{(NL)}(0^+) =
1$.} \label{fig3}
\end{figure}

To illustrate our self-consistent theoretical framework, we solve
Eq.\ (\ref{beq}) numerically for $\tau/\tau_0 \rightarrow 0$,
and plot the resulting temporal autocorrelation
function of diffusely reflected wave, $g_1^{(NL)}(0^+)$, in Fig.\ \ref{fig3}.
As discussed above, $g_1^{(NL)}(0^+) < 1$ corresponds to spontaneous
fluctuations of scattered waves which is a manifestation of
the speckle pattern instability.
It follows from Fig.\ \ref{fig3}, that in accordance with
Eq.\ (\ref{stbl}), an
infinitely small $\left| \Delta n \right|$ is sufficient to make the
speckle pattern unstable
in the absence of absorption, while a certain threshold degree of nonlinearity
is required to destabilize the speckle pattern
in a dissipative medium.
In the absence of absorption, Eq.\ (\ref{beq}) always has two solutions:
$\beta(0^+) = 0$ and $\beta(0^+) > 0$, corresponding to
$g_1^{(NL)}(0^+) = 1$ and $g_1^{(NL)}(0^+) < 1$, respectively.
As discussed above, it is the second solution, shown by a dashed line in
Fig.\ \ref{fig3}, which is the physically realizable one.
In contrast, if $\ell/L_a \neq 0$, $\left| \Delta n \right|$ should be
greater than some threshold value
for the second solution $\beta(0^+) > 1$ to appear.
Threshold values of $\left| \Delta n \right|$ following from
Eq.\ (\ref{stbl}) are shown in Fig.\ \ref{fig3} by arrows.

For a nonabsorbing, elastically scattering medium ($\ell/L_a = 0$),
the value of $g_1^{(NL)}(0^+)$ can be estimated analytically.
Indeed, at $\tau/\tau_0 \rightarrow 0$ the principal contribution to
$\left< \Delta\varphi^2(\tau) \right>_s$ is given
by $\left< \Delta\varphi^2(\tau) \right>_s^{(2)}$
for $s/\ell \ll (k_0 \ell)^2$, and by
$\left< \Delta\varphi^2(\tau) \right>_s^{(3)}$
for $s/\ell \gg (k_0 \ell)^2$.
This allows us to put
$\left< \Delta\varphi^2(\tau) \right>_s \approx
\left< \Delta\varphi^2(\tau) \right>_s^{(2)}$ if
$s/\ell < (k_0 \ell)^2$, and
$\left< \Delta\varphi^2(\tau) \right>_s \approx
\left< \Delta\varphi^2(\tau) \right>_s^{(3)}$
if $s/\ell > (k_0 \ell)^2$.
Integration in Eq.\ (\ref{pathnl}) can be then carried out,
and Eq.\ (\ref{beq}) is easily solved, yielding
\begin{eqnarray}
&&g_1^{(NL)}(0^+) \nonumber \\
&&\simeq 1-\cases{ 2 \left| \Delta n \right|^{2/3}, &$\left| \Delta
n \right| < (k_0 \ell)^{-3/2}$, \cr 3 \left| \Delta n \right|
\sqrt{k_0 \ell}, &$\left| \Delta n \right| > (k_0 \ell)^{-3/2}$.}
\label{b0}
\end{eqnarray}
This result agrees well with the numerical solution of Eq.\ (\ref{beq})
presented in Fig\ \ref{fig3}.

The physical origin of the instability of speckle pattern for $p>1$
can be revealed by realizing that the system ``coherent wave + nonlinear
disordered medium'' has a positive three-dimensional
feedback.
In a nonlinear medium,
an infinitely small perturbation of the wave intensity $I(\vec{r}, t)$
produces a change
of the local refractive index, which alters the phases of waves
propagating in the medium and, consequently, affects their
mutual interference. Since it is this interference which determines
$I(\vec{r}, t)$, the loop of the feedback is closed.
For $p \gtrsim 1$, the feedback is sufficiently strong
to compensate for the (diffusive on average) spreading of the
initial intensity perturbation, and the speckle pattern $I(\vec{r}, t)$
is unstable.
It is worthwhile to note that unstable regimes are not exceptional
in nonlinear wave systems and, in particular, in optical
systems (see, e.g.,
Refs.\ \onlinecite{gibbs85,vor95,aless91,arecchi99,ramazza96}).

\begin{figure}[t]
\vspace{0.5cm} \psfig{file=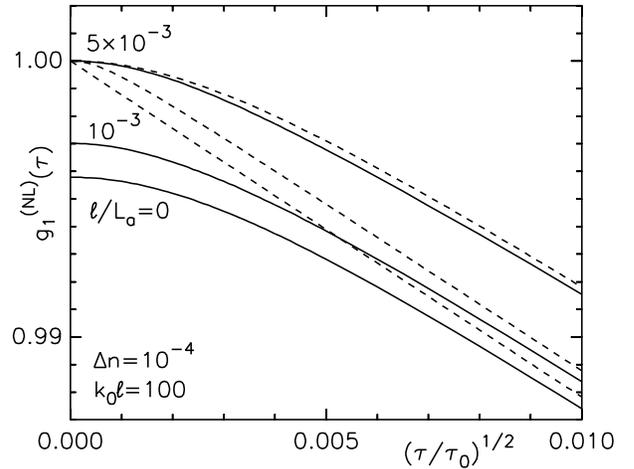,width=8cm} \vspace{0.5cm}
\caption{The same as Fig.\ \ref{fig1}, but using a self-consistent
approach instead of the perturbation theory. For the two lower
curves, $g_1^{(NL)}(0^+) < 1$ and the speckle pattern is unstable.}
\label{fig4}
\end{figure}

Our Eq.\ (\ref{beq}) is in no way limited to the case of $\tau/\tau_0
\rightarrow 0$, and can be used to compute the temporal autocorrelation
function of diffusely reflected wave at any $\tau/\tau_0 > 0$.
In the latter case, one should take into account all
the four dephasing terms given by Eqs.\ (\ref{dphi0}),
(\ref{d11})--(\ref{d33}).
The results of the numerical solution of Eq.\ (\ref{beq}) are
shown in Fig.\ \ref{fig4}.
For weak absorption ($\ell/L_a = 0, 10^{-3}$),
$p > 1$ and the speckle pattern is unstable [$g_1^{(NL)}(0^+) < 1$].
As $\tau/\tau_0$ increases, the difference between
the ``nonlinear'' and ``linear'' curves becomes less pronounced.
For strong absorption ($\ell/L_a = 5 \times 10^{-3}$),
$p$ becomes less than unity
and stability of the speckle pattern is recovered
[$g_1^{(NL)}(0^+) = 1$].
We remind that the temporal autocorrelation function $g_1^{(L)}(\tau)$,
corresponding to a linear medium, always equals $1$ for
$\tau/\tau_0 = 0$, as shown by dashed lines in Fig.\ \ref{fig4}.

\begin{figure}
\vspace{0.5cm} \psfig{file=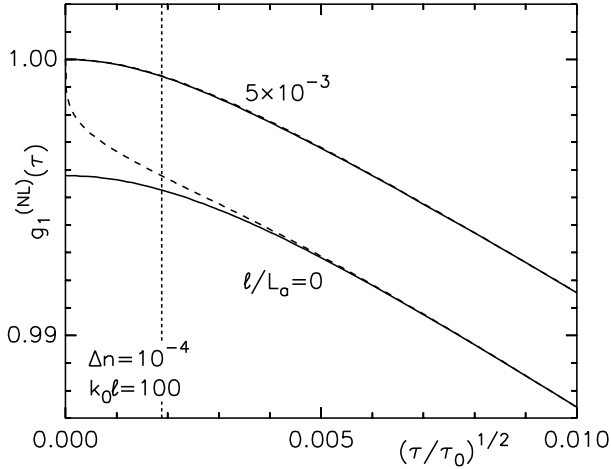,width=8cm} \vspace{0.5cm}
\caption{Comparison of the normalized temporal autocorrelation
functions calculated using a self-consistent (solid lines) and
perturbation (dashed lines) theories. Dotted vertical line
indicates the limit $\tau_c$ of validity of the perturbation theory
for $\ell/L_a = 0$. The perturbation theory is valid only for $\tau
> \tau_c$, while for $\tau < \tau_c$ the perturbation result
deviates significantly from the self-consistent solution. For
sufficiently strong absorption ($\ell/L_a = 5 \times 10^{-3}$), the
perturbation theory holds at any $\tau$: the perturbation and the
self-consistent results are indistinguishable.} \label{fig5}
\end{figure}

It is instructive to compare the results obtained
using the perturbation theory of Sec.\ \ref{perturb} and the
self-consistent approach of Sec.\ \ref{self}.
Such a comparison is shown in Fig.\ \ref{fig5}.
The two upper curves,
corresponding to relatively strong absorption
($\ell/L_a = 5 \times 10^{-3}$, $p < 1$),
are almost indistinguishable, which means that for $p<1$,
the perturbation theory works very well.
In contrast, for $p>1$ (see the two lower curves of Fig.\ \ref{fig5},
corresponding to $\ell/L_a = 0$),
the perturbation and the self-consistent curves are close only
at the right from the dotted vertical line, showing the minimum
time $\tau_c$ at which the perturbation theory is valid
[see Eq.\ (\ref{validity})]. For $\tau \lesssim \tau_c$, the perturbation
and the self-consistent results disagree significantly,
which confirms our conclusion about the failure of the perturbation
approach at short correlation delay times.

\section{Conclusion}
\label{concl}

We are now in a position to answer the two central questions
formulated in the introductory section:
\begin{itemize}

\item[({\em a\/}).]
The phenomenon of multiple scattering is capable of providing
a positive feedback for a coherent wave propagating in a nonlinear
disordered medium.

\item[({\em b\/}).]
The onset of the speckle pattern instability
occurs when the control parameter
\begin{eqnarray}
p =
\Delta n^2 \left(\frac{L_a}{\ell} \right)^2
\left[ k_0 \ell + \frac{L_a}{\ell} \right]
\label{stbl1}
\end{eqnarray}
becomes of order or larger than unity. The speckle pattern
is stable for $p < 1$.

\end{itemize}
The instability of the multiple-scattering
speckle pattern manifests itself in spontaneous fluctuations of
the scattered wave field and intensity. The following features
are characteristic for the development of the instability.
First, the development of the instability
is independent of the sign of nonlinearity.
This is not common for nonlinear waves since the instability is
often due to self-focusing phenomena, which only occur
for $n_2 > 0$.\cite{gibbs85,vor95,aless91,arecchi99,ramazza96}
The instability of waves in a {\em disordered\/}
nonlinear medium has nothing to do with the self-focusing,
and occurs at relatively weak nonlinearities, when the self-focusing
can be neglected.
Second, in the absence of absorption, the speckle pattern is unstable for
any (even infinitely small) value of the nonlinearity strength
$\left| \Delta n \right|$,
while in a dissipative medium $\left| \Delta n \right|$
should exceed a certain threshold value for the instability
to show up.
Finally, the instability results in a value of the
autocorrelation function $g_1^{(NL)}(\tau)$ of scattered wave,
which is smaller than $1$ for $\tau/\tau_0 \rightarrow 0$
(i.e., in the absence of scatterer motion).
This is a clear signature of spontaneous fluctuations of the
multiple-scattered speckle pattern, and should be observable
in experiments.

\acknowledgements
The author is indebted to R. Maynard for numerous discussions and
continuous support of this work.

\appendix
\section{Spatio-temporal filed correlation in a linear medium}
\label{appa}

In this appendix, we provide a derivation of the spatio-temporal
correlation function,
$C_{\psi}(\vec{r}, \Delta\vec{r}, \tau) =
\left< \psi(\vec{r}-\Delta\vec{r}/2, t)
\psi^*(\vec{r}+\Delta\vec{r}/2, t+\tau) \right>$,
of a random field
$\psi(\vec{r}, t)$ in the bulk of disordered medium.
Starting from the linear wave equation [Eq.\ (\ref{weq}) with
$\varepsilon_2 = 0$], we obtain the Bethe-Salpeter equation
in the form:\cite{pnini89,frisch68,shapiro86}
\begin{eqnarray}
&&C_{\psi}(\vec{r}, \Delta\vec{r}, \tau) = \left<
\psi(\vec{r}-\Delta\vec{r}/2, t) \right> \left<
\psi^*(\vec{r}+\Delta\vec{r}/2, t) \right> \nonumber \\
&&+ \int d \vec{r}_a \int d \vec{r}_b \int d \vec{r}_c \int d
\vec{r}_d \overline{G}(\vec{r}-\Delta\vec{r}/2, \vec{r}_a)
\nonumber \\
&&\times\, \overline{G}^*(\vec{r}+\Delta\vec{r}/2, \vec{r}_b)
U(\vec{r}_a, \vec{r}_b, \vec{r}_c, \vec{r}_d) \nonumber \\
&&\times\, C_{\psi}[(\vec{r}_c + \vec{r}_d)/2, \vec{r}_d -
\vec{r}_c, \tau], \label{bethe}
\end{eqnarray}
where the integrations are over the volume of disordered medium,
$\overline{G}$ is the average Green function of the linear wave equation,
and $U$ is the irreducible four-point vertex.
Far enough from the medium boundaries,
we can replace $\overline{G}$ by its value in the infinite medium:
\begin{eqnarray}
&&\overline{G}(\vec{r}_1, \vec{r}_2) \nonumber \\
&&= -\frac{1}{4 \pi \left| \vec{r}_1 - \vec{r}_2 \right|}
\exp\left[\left(i k -\frac{1}{2 \ell} \right) \left| \vec{r}_1 -
\vec{r}_2 \right| \right]. \label{grfunc}
\end{eqnarray}

Now we assume that point scatterers in the medium undergo Brownian motion,
and that the correlation function of the dielectric function
fluctuations is given by Eq.\ (\ref{cde}).
For $\tau \ll \tau_0$, we can neglect the $\tau$-dependence of
$U$ in Eq.\ (\ref{bethe}), which in the limit of $k \ell \gg 1$ becomes
\begin{eqnarray}
&&U(\vec{r}_a, \vec{r}_b, \vec{r}_c, \vec{r}_d) \nonumber \\
&&= \frac{4 \pi}{\ell} \delta(\vec{r}_a - \vec{r}_c)
\delta(\vec{r}_b - \vec{r}_d) \delta(\vec{r}_a - \vec{r}_b).
\label{u}
\end{eqnarray}
Eq.\ (\ref{bethe}) then reduces to
\begin{eqnarray}
&&C_{\psi}(\vec{r}, \Delta\vec{r}, \tau) = \frac{4 \pi}{\ell} \int
d \vec{r}_a\, \overline{G}(\vec{r}-\Delta\vec{r}/2, \vec{r}_a)
\nonumber \\
&&\times\, \overline{G}^*(\vec{r}+\Delta\vec{r}/2, \vec{r}_a)
C_{\psi}(\vec{r}_a, 0, \tau), \label{bethe2}
\end{eqnarray}
where we assumed the coherent field
$\left< \psi(\vec{r}, t) \right>$ to be negligible.

Now we remind that $C_{\psi}(\vec{r}_a, 0, \tau)
\equiv C_{\psi}(\vec{r}_a, \tau)$
is a slow-varying function of $\vec{r}_a$. We can therefore
pull it out of the integral in Eq.\ (\ref{bethe2}), taking its
value at $\vec{r}$.
Performing the remaining integral, we find
\begin{eqnarray}
C_{\psi}(\vec{r}, \Delta\vec{r}, \tau) =
\frac{\sin(k \Delta r)}{k \Delta r}
\exp\left( -\frac{\Delta r}{2 \ell} \right)
C_{\psi}(\vec{r}, \tau).
\label{cf1}
\end{eqnarray}
This equation relates the spatio-temporal correlation of
the field with its purely temporal correlation. Eq.\ (\ref{cf1})
holds for any sample geometry and source distribution,
far enough from boundaries and sources, and for $\tau \ll \tau_0$.
In the case of a plane wave incident at the surface $z = 0$ of
a semi-infinite disordered medium, occupying the $z>0$ half-space,
\begin{eqnarray}
C_{\psi}(\vec{r}, \tau) =
I_0 \exp\left[ -\alpha(\tau) (z/\ell) \right]
\label{cf1t}
\end{eqnarray}
with $\alpha^2(\tau) = 3 \tau/(2 \tau_0) + (\ell/L_a)^2$,
leading to Eq.\ (\ref{tfield}) in the main text.
In the factorization approximation (\ref{tint1}), the square
of Eq.\ (\ref{cf1}) gives the short-range correlation function
of intensity fluctuations.

\section{Long-range spatio-temporal correlation of intensity
fluctuations in a linear medium}
\label{appb}

To calculate the spatio-temporal correlation function of intensity fluctuations,
$C_{\delta I}(\vec{r}, \Delta\vec{r}, \tau) =
\left< \delta I(\vec{r}-\Delta\vec{r}/2, t)
\delta I(\vec{r}+\Delta\vec{r}/2, t+\tau) \right>$,
for $\Delta r > \ell$
we generalize the Langevin approach,\cite{zyuzin87,pnini89,lh} which
has been initially developed for calculation of purely spatial correlations.
We assume that the spatio-temporal correlation function of Langevin random
sources is determined by the short-range correlation function of
intensity fluctuations. In the factorization approximation,
the latter is given by the square of the field-field correlation
function [Eq.\ (\ref{cf1})]. Hence, the generalized Langevin equations read
\begin{eqnarray}
&&D_p \left[ \nabla^2 - 1/L_a^2 \right] \delta I(\vec{r}, t) = {\rm
div} \vec{j}_{ext}(\vec{r}, t), \label{lgvn1}
\\
&&\left< j_{ext}^{(i)}(\vec{r}-\Delta\vec{r}/2, t)\,
j_{ext}^{(m)}(\vec{r}+\Delta\vec{r}/2, t+\tau) \right> \nonumber
\\
&&= \frac{1}{3} \delta_{i m} \frac{c^2 2\pi\ell}{k^2} \left|
C_{\psi}(\vec{r}, \tau) \right|^2 \delta(\Delta\vec{r}),
\label{lgvn2}
\end{eqnarray}
where $D_p = c \ell/3$ is the diffusion coefficient, and $c$ is the
speed of wave in the medium.
In the case of a plane wave incident upon a boundary $z = 0$
of a semi-infinite disordered
medium, it is convenient to make a two-dimensional Fourier transform
of Eq.\ (\ref{lgvn1}) in the $\left\{ x, y \right\}$ plane.
Then the equations corresponding to
$\delta I(\vec{K}_1, z_1, t_1)$ and $\delta I(\vec{K}_2, z_2, t_2)$
are multiplied, and the product is ensemble averaged.
Further transformations, which are equivalent to those discussed in
Ref.\ \onlinecite{pnini89}, yield
\begin{eqnarray}
&&\left< \delta I(\vec{K}_1, z_1, t_1) \delta I(\vec{K}_2, z_2,
t_2) \right> = I_0^2 \frac{6 \pi}{k^2 \ell} \delta(\vec{K}_1 -
\vec{K}_2) \nonumber \\
&&\int_0^{\infty} d z^{\prime} \Big[ (\vec{K_1} \cdot \vec{K}_2)
G(p, z_1, z^{\prime}) G(p, z_2, z^{\prime})
\nonumber \\
&&+\, \left.\frac{\partial}{\partial z^{\prime}} G(p, z_1,
z^{\prime}) \frac{\partial}{\partial z^{\prime}} G(p, z_2,
z^{\prime}) \right] \exp\left[-2 \alpha(\tau)
\frac{z^{\prime}}{\ell} \right], \label{ik}
\end{eqnarray}
where $p^2 = \vec{K}^2 + 1/L_a^2$, and
\begin{eqnarray}
G(p, z_1, z^{\prime}) &=& -\frac{1}{p} \sinh(p \times
\min\left\{z_1,
z^{\prime}\right\}) \nonumber \\
&\times& \exp(-p \times \max\left\{z_1, z^{\prime}\right\})
\label{gr}
\end{eqnarray}
is the Green function of Eq.\ (\ref{lgvn1}).
Evaluating Eq.\ (\ref{ik}) and transforming the result back to the
real space, after lengthy but straightforward algebra we obtain
\begin{eqnarray}
C_{\delta I}(\vec{r}, \Delta\vec{r}, \tau) &=& \frac{3}{(k \ell)^2}
I_0^2 \int_0^{\infty} d K\, K \nonumber \\
&\times& Q \left[ K, \sqrt{K^2 + \ell^2/L_a^2}, \frac{z}{\ell},
\frac{\Delta z}{\ell}, \alpha(\tau) \right] \nonumber \\
&\times& J_0\left( K \frac{\Delta R}{\ell} \right), \label{ir}
\end{eqnarray}
where
\begin{eqnarray}
&&Q(K, p, \zeta, \Delta \zeta, \alpha) =
\exp\left( 2 p \zeta \right) \times
\left[ \frac{2 \alpha^2 + K^2 - p^2}{4 \alpha^3 - 4 \alpha p^2} \right.
\nonumber \\
&&\left.-\left[(\alpha^2 - p^2)(-K^2 + p^2) + \alpha^2 (K^2 + p^2)
\cosh(2 p \zeta_1) \right.\right. \nonumber \\
&&+ \left.\left. \alpha p (K^2 + p^2) \sinh(2 p \zeta_1)
\right]/\left[4 \alpha \exp(2 \alpha \zeta_1)(\alpha^2 - p^2) p^2
\right] \right]
\nonumber \\
&&+\left( 1 + \frac{K^2}{p^2} \right) \frac{ \sinh(p \zeta_1) \sinh
(p \zeta_2)}{2 \exp\left[ 2 (\alpha + p) \zeta_2 \right] (\alpha +
p)} \nonumber \\
&&+\frac{\exp(-p \zeta_2) \sinh(p \zeta_1)}{4 \alpha p^2
(\alpha+p)}
\nonumber \\
&&\times \left\{ p(p^2 + 2\alpha p - K^2) \left[\cosh(p \zeta_2)
\exp[-(2\alpha + p) \zeta_2] \right.\right.\nonumber \\
&&- \left.\left. \cosh(p \zeta_1) \exp[-(2\alpha + p) \zeta_1]
\right] \right.
\nonumber \\
&&\left.+(p^3 - 2\alpha K^2 - pK^2 ) \left[\sinh(p \zeta_2)
\exp[-(2\alpha + p) \zeta_2] \right.\right.\nonumber \\
&&\left.\left.- \sinh(p \zeta_1) \exp[-(2\alpha + p) \zeta_1]
\right] \right\}, \label{q}
\end{eqnarray}
where $\zeta_1 = \zeta-\Delta\zeta/2$ and
$\zeta_2 = \zeta+\Delta\zeta/2$.

Equations (\ref{ir}), (\ref{q}) cannot be evaluated in a general
form. In the limits of
$\alpha(\tau) \zeta, \alpha(\tau) \Delta R/\ell
\ll 1$ we can, however, approximately replace $Q$ in Eq.\ (\ref{ir}) by
\begin{eqnarray}
Q(K, p, \zeta, \Delta \zeta, \alpha) &\approx& \frac{1}{2 K} \left[
\exp(-K \Delta\zeta) - \exp(-2 K \zeta) \right] \nonumber \\
&\times& \exp[-2 \alpha(\tau) \zeta], \label{q0}
\end{eqnarray}
which yields Eq.\ (\ref{lrs}) in the main text.

\section{Density of wave paths in a disordered medium}
\label{appc}

Propagation of waves in a disordered medium can
be interpreted in terms of partial waves traveling along
various paths inside the medium. The spatial distribution
of such paths and their relative weights depend on the scattering
properties of the medium, and on the geometry of the sample.
In the case of multiple scattering, the simplest and, at the same
time, sufficiently accurate model of wave propagation is the
{\em diffusion\/} model. According to this model, wave paths
in the medium coincide with trajectories of Brownian particles.
The probability $G(\vec{r}_1, \vec{r}_2, s)$
for a path of length $s$ to pass from
$\vec{r}_1$ to $\vec{r}_2$ is then given by a solution of
the diffusion equation, which in the absence of absorption
reads:\cite{bicout93,akker88}
\begin{eqnarray}
\frac{d}{d s}G(\vec{r}_1, \vec{r}_2, s) - \frac{1}{3} \ell \nabla^2
G(\vec{r}_1, \vec{r}_2, s) = \delta(\vec{r}_1 - \vec{r}_2)
\delta(s), \label{de}
\end{eqnarray}
where $\ell$ is the mean free path.
Commonly used
boundary conditions for Eq.\ (\ref{de}) consist in putting
$G = 0$ at open boundaries and $\nabla_n G = 0$
at reflecting boundaries of the sample (where $\nabla_n$
denotes the normal derivative of $G$).
$G(\vec{r}_1, \vec{r}_2, s)$ is called the Green function, or the
propagator.
For a semi-infinite medium occupying the half-space $z>0$,
one finds
\begin{eqnarray}
G(\vec{r}_1, \vec{r}_2, s) &=& \left( \frac{3}{4 \pi \ell s}
\right)^{3/2} \left\{ \exp\left[-\frac{3}{4 \ell s} \left(\Delta
R^2 + \Delta z^2 \right) \right] \right.\nonumber \\
&-& \left. \exp\left[-\frac{3}{4 \ell s} \left(\Delta R^2 + Z^2
\right) \right] \right\}, \label{green}
\end{eqnarray}
where cylindrical coordinates are used:
$\vec{r}_i = \left\{ \vec{R}_i, z_i \right\}$,
$\Delta \vec{R} = \vec{R}_1 - \vec{R}_2$,
$\Delta z = z_1 - z_2$, and $Z = z_1 + z_2$.

Following Ref.\ \onlinecite{bicout93}, we introduce
$\rho_s(\vec{r}_1, \vec{r}_2, \vec{r}_3)$, the {\em density
distribution\/} of paths of length $s$, as a number of visits of a
given site $\vec{r}_2$ inside $d^3 \vec{r}_2$ in the ensemble of
paths of length $s$ starting at $\vec{r}_1$ and ending at
$\vec{r}_3$, over the total length of the ensemble distinct paths:
\begin{eqnarray}
\rho_s(\vec{r}_1, \vec{r}_2, \vec{r}_3) &=& \frac{1}{s G(\vec{r}_1,
\vec{r}_3, s)} \int_0^s d p\, G(\vec{r}_1, \vec{r_2}, p) \nonumber
\\ &\times& G(\vec{r}_2, \vec{r}_3, s-p). \label{rhoadef}
\end{eqnarray}
$\rho_s(\vec{r}_1, \vec{r}_2, \vec{r}_3)$ describes the probability density
for a path of a given length $s$,
starting at $\vec{r}_1$ and ending at $\vec{r}_3$, to
pass through $\vec{r}_2$. This quantity is normalized:
\begin{eqnarray}
\int d^3 \vec{r}_2\, \rho_s(\vec{r}_1, \vec{r}_2, \vec{r}_3) = 1,
\label{rhoanorm}
\end{eqnarray}
where the integration is performed over the volume of disordered medium.

As the Green function $G$ is known [Eq.\ (\ref{green})], the calculation
of $\rho_s(\vec{r}_1, \vec{r}_2, \vec{r}_3)$ is straightforward.
For diffusely reflected paths,
assuming that the first and the last scattering events take place
at $z = \ell$, we obtain
\begin{eqnarray}
\rho_s(\ell, \vec{r}, \ell) \equiv \rho_s(\vec{r}) =
\frac{1}{A}\frac{6 z}{\ell s} \exp\left( -\frac{3 z^2}{\ell s} \right),
\label{rhoa}
\end{eqnarray}
where $A \rightarrow \infty$ is the surface of the semi-infinite
medium.
Equation (\ref{rhoa}) defines the probability density for a diffusely
reflected path of length $s$ to pass through a vicinity of
some point $\vec{r} = \{x, y, z \}$.

Generalizing definition of $\rho_s$, we define the probability density
for a path of length $s$
starting at $\vec{r}_1$ and ending at $\vec{r}_4$ to
pass consequently through $\vec{r}_2$ and $\vec{r}_3$:
\begin{eqnarray}
\rho_s(\vec{r}_1, \vec{r}_2, \vec{r}_3, \vec{r}_4) &=& \frac{2}{s^2
G(\vec{r}_1, \vec{r}_4, s)} \int_0^s d p \int_0^{s-p} d q
\nonumber \\
&&G(\vec{r}_1, \vec{r_2}, p) G(\vec{r}_2, \vec{r}_3, q) \nonumber
\\ &\times&
G(\vec{r}_3, \vec{r}_4, s-p-q). \label{rhobdef}
\end{eqnarray}
The normalization of Eq.\ (\ref{rhobdef}) is
\begin{eqnarray}
\int d^3 \vec{r}_2 \int d^3 \vec{r}_3\, \rho_s(\vec{r}_1,
\vec{r}_2, \vec{r}_3, \vec{r}_4) = 1. \label{rhobnorm}
\end{eqnarray}
For a semi-infinite medium, we get
\begin{eqnarray}
\rho_s(\ell, \vec{r}, \vec{r}^{\prime}, \ell) &\equiv&
\rho_s(\vec{r}, \vec{r}^{\prime}) =
\frac{1}{A^2} \frac{9}{2 \pi \ell^2 s^2}
\nonumber \\
&\times&\left\{\frac{Z + \sqrt{\Delta R^2 + \Delta
z^2}}{\sqrt{\Delta R^2 + \Delta z^2}} \right. \nonumber
\\
&\times& \left.\exp\left[ -\frac{3}{4 \ell s}\left(Z + \sqrt{\Delta
R^2 + \Delta z^2} \right)^2 \right] \right. \nonumber
\\
&-& \left.\frac{Z + \sqrt{\Delta R^2 + Z^2}}{\sqrt{\Delta R^2 +
Z^2}} \right. \nonumber \\
&\times& \left.\exp\left[ -\frac{3}{4 \ell s}\left(Z + \sqrt{\Delta
R^2 + Z^2} \right)^2 \right] \right\}, \label{rhob}
\end{eqnarray}
where $\vec{r} = \left\{ \vec{R}, z \right\}$,
$\Delta \vec{R} = \vec{R} - \vec{R}^{\prime}$,
$\Delta z = z - z^{\prime}$, $Z = z + z^{\prime}$.

Although Eqs.\ (\ref{rhoa}) and (\ref{rhob}) have been found
for a nonabsorbing medium, it is easy to show that these
results hold in the presence of {\em spatially-homogeneous\/}
absorption as well. This stems from the fact that the attenuation
of wave in a homogeneously absorbing medium depends only on the
path {\em length\/}, while being independent of the path
{\em geometry.} As a consequence, the Green function [Eq.\ (\ref{green})]
should be multiplied by a factor
$\exp(-s/\ell_a)$, where $\ell_a$ is the absorption length.
This factor, however, disappears after the substitution of
the Green function (\ref{green}) in Eqs.\ (\ref{rhoadef}) and
(\ref{rhobdef}). Consequently, $\rho_s(\vec{r})$ and
$\rho_s(\vec{r}, \vec{r}^{\prime})$ are
independent of $\ell_a$ and remain unchanged.


\begin{references}
\bibitem{optics}
For the sake of concreteness, we use the ``optical'' language throughout
the paper. Our results are, however, general, and apply to all types
of scalar waves (in the case of optical waves, polarization effects
are neglected).
\bibitem{unstable1}
Although we take the limit of $\tau/\tau_0 \rightarrow 0$, we
assume $\tau \gg T_{jump}$, where $T_{jump} \gg 2\pi/\omega$ is the
typical time required for the system to ``jump'' from one solution
of Eq.\ (\ref{weq}) to another. The time autocorrelation function
$g_1^{(NL)}(\tau)$ is independent of $\tau$ for $\tau \gg T_{jump}$
and $\tau/\tau_0 \rightarrow 0$.\cite{skip00b}
\bibitem{lh}
One can check, however, that exactly the same result follows
from a diagrammatic calculation where interaction of four
ladders is described with help of the Hikami vertex.
\end{references}
\end{document}